\documentclass[
a4paper,twocolumn,
superscriptaddress, 
amsmath,amssymb,
]{quantumarticle}
\pdfoutput=1

\usepackage{graphicx}% Include figure files
\usepackage{subcaption}
\usepackage{listings}
\usepackage{dcolumn}% Align table columns on decimal point
\usepackage{bm}% bold math
\usepackage{hyperref}% add hypertext capabilities
\usepackage[english]{babel}
\usepackage[T1]{fontenc}
\hypersetup{
    colorlinks=true,
    linkcolor=blue,
    filecolor=blue,      
    urlcolor=blue,
    pdftitle={Exploring polymer classification with a hybrid single-photon quantum approach},
    pdfpagemode=FullScreen,
    citecolor=blue,
    }
\usepackage[utf8]{inputenc}
\usepackage{booktabs, multirow}
\usepackage[numbers,sort&compress]{natbib}
\begin{document}

\title{Exploring polymer classification with a hybrid single-photon quantum approach} % Force line breaks with \\

\author{Alexandrina Stoyanova and Bogdan Penkovsky}%
\affiliation{%
 Alysophil \\
 850 Bd Sébastien Brant, 67400 Illkirch-Graffenstaden, France 
}%

\begin{abstract}
Polymers exhibit complex architectures and diverse properties that place them at the center of contemporary research in chemistry and materials science.~As conventional computational techniques, even multi-scale ones, struggle to capture this complexity, quantum computing offers a promising alternative framework for extracting structure-property relationships.~Noisy Intermediate-Scale Quantum (NISQ) devices are commonly used to explore the implementation of algorithms, including quantum neural networks for classification tasks, despite ongoing debate regarding their practical impact.

We present a hybrid classical–quantum formalism that couples a classical deep neural network for polymer featurization with a single-photon-based quantum classifier native to photonic quantum computing.~This pipeline successfully classifies polymer species by their optical gap, with performance in line between CPU-based noisy simulations and a proof-of-principle run on Quandela’s Ascella quantum processor.~These findings demonstrate the effectiveness of the proposed computational workflow and indicate that chemistry-related classification tasks can already be tackled under the constraints of today’s NISQ devices.
\end{abstract}

\maketitle

\section{\label{sec:intro}Introduction}
During the last decade emerging quantum algorithms have been proposed to simulate complex chemical structures and reactions \cite{ref_QC_emulation, ref_Grid_based, ref_QC_advantage, Mazzola2023}.~Their development was boosted by the rapid progress of computational devices that harness quantum phenomena and thus serve as platforms for performing quantum calculations.~Various physical systems have been intensively studied as carriers of quantum information such as superconducting circuits \cite{ref_supercondc_processor}, trapped ions \cite{PhysRevX.13.041052}, neutral atoms \cite{ref_quantum_processor_atoms} and photons \cite{PhysRevLett.127.180502, quantum_circuits_photons, Ascella_2023, ref_quantum_advantage_photons,ref_quantum_advantage_photons_2}.~Early demonstrations of quantum computational advantage in sampling tasks have already been achieved using superconducting \cite{ref_supercondc_processor} and photon-based \cite{ref_quantum_advantage_photons, ref_quantum_advantage_photons_2} devices. 

Quantum algorithms are particularly poised to bring significant advances in chemistry, materials science, and bio-sciences, where the intrinsic complexity of macromolecules, polymers, and composites generates vast chemical spaces.~Exploring the entire chemical space of such complex structures is computationally intractable, even for advanced classical or quantum algorithms running on conventional hardware.~Instead, only limited portions of this vast chemical space are explored
when searching for structures with specific properties (see, e.~g., Refs.~\cite{ref1, QC_polymers, data_driven, ref5, ref7, ref30, ref33, ref3}).~Existing models and methods (as reviewed in Refs.~\cite{ref1} and~\cite{kunchapu2025polymetrix})
operate independently at various length and time scales, aligning with the chemical and topological structures and dynamics of polymers across a wide spectrum of scales \cite{ref_multi_scale, ref1, ref2, data_driven, ref3, LM1}.~These approaches are justified because the structure and dynamics at length scales comparable to the polymer's chain length might be unaffected by the monomer's chemical composition (atomic types and bonding), and vice versa.~When the studied properties are intrinsic to the monomer’s chemistry like the band gap, electronic polarizability, and dielectric constant, the
atomic ($\sim$10$^{-9}$ m, $\sim$10$^{-9}$–10$^{-6}$ s) and even quantum scales ($\sim$10$^{-10}$ m, $\sim$10$^{-12}$ s) are involved.~Then, a quantum mechanical treatment can be applied to the monomer (see Review \cite{QC_polymers}), but computational costs become quickly prohibitive with the size and complexity of the monomer, even in classical devices endowed with Graphic Processing Units (GPUs), and even by using share-memory and parallelization strategies.~The computational challenge is even greater when different scales need to be treated on equal footing in polymer property description \cite{ref_multi_scale, ref1, ref23, ref31, ref32}.

In parallel to conventional quantum chemical \cite{QC_polymers}, atomistic and coarse-grained molecular dynamics \cite{ref3, ref23, ref1, ref33} approaches to polymers, data-driven Machine and Deep Learning (ML and DL) dedicated algorithms \cite{data_driven, ref4, ref5, ref6, ref7, ref8, ref9, ref10, ref11, ref12, ref23, ref28, ref29, ref30, ref31, ref32, kunchapu2025polymetrix} have flourished the last decade.~Key ingredients in this approach are property- and structure-related data, along with machine-readable polymer representations, known as features.~Several featurization schemes have been proposed to transform the hierarchical and stochastic polymer structures into chemically-informed, machine-readable representations \cite{ref9, ref21, ref13, ref29, ref_genBigSMILES, ref_canonBigSMILES, ref_covalBigSMILES}.~These featurization schemes may operate at various length scales depending on the properties under study, ranging from more complex representations for ensembles of macromolecules or macromolecular chains to simpler ones for monomeric chemical structures.~Similar to the conventional methods, the structure-property correlation patterns in polymer data, described by complex featurization schemes, can be challenging to detect using classical machines running traditional machine learning (ML) and deep learning (DL) algorithms.~Quantum computing, in contrast, enables calculations in higher-dimensional spaces, potentially allowing for more sophisticated learning models and complex feature representations.

Yet, quantum computations for macromolecules or their stochastic mixtures with intricate chemical and topological structures would require a substantial number of high-quality, low-decoherence qubits.~This holds especially for highly accurate quantum chemical molecular or material studies.~Even at small length scales comparable to monomer size, the intrinsic structural complexity of monomeric units could impede quantum computations unless suitable conceptual or computational reformulations (e.g., coarse-graining, effective models, or problem-specific encodings) are employed.~This is so, because even a single monomer has many degrees of freedom - multiple atoms, electronic orbitals, vibrational modes, and conformational states that increase the size of the Hilbert space.~Electronic, vibrational, and geometric degrees of freedom are also often strongly coupled that prevents simple separation of variables in algorithms.~Accurately encoding such complex Hamiltonians would require a large number of qubits and deep quantum circuits.~Furthermore, in quantum ML (QML) and DL approaches, encoding the polymer data into a machine-readable format may require a substantial number of logical qubits, with the exact count depending on the chosen featurization scheme.~Despite the challenges in implementing quantum algorithms for comprehensive simulations of complex chemical systems, which would require many error-corrected qubits, promising results may still be achievable with near-term quantum approaches \cite{QuantumChemistryQubits, QuantumChemistryQubits2, materials_on_NISQ}. 

Photonics presents a promising platform for quantum computing.~Besides the near-term demonstrations of quantum computational advantage \cite{ref_quantum_advantage_photons_2}, photonic systems possess valuable properties, such as low decoherence rates and the ability to encode information across multiple degrees of freedom.~In the context of near-term algorithms, particularly QML, photonic-native models leveraging Fock-space encoding can be implemented, potentially leading to enhanced expressivity of the model with fixed computational resources \cite{ref42}.

\begin{figure*}[!htb]
    \centering    \includegraphics[width=1.0\textwidth]{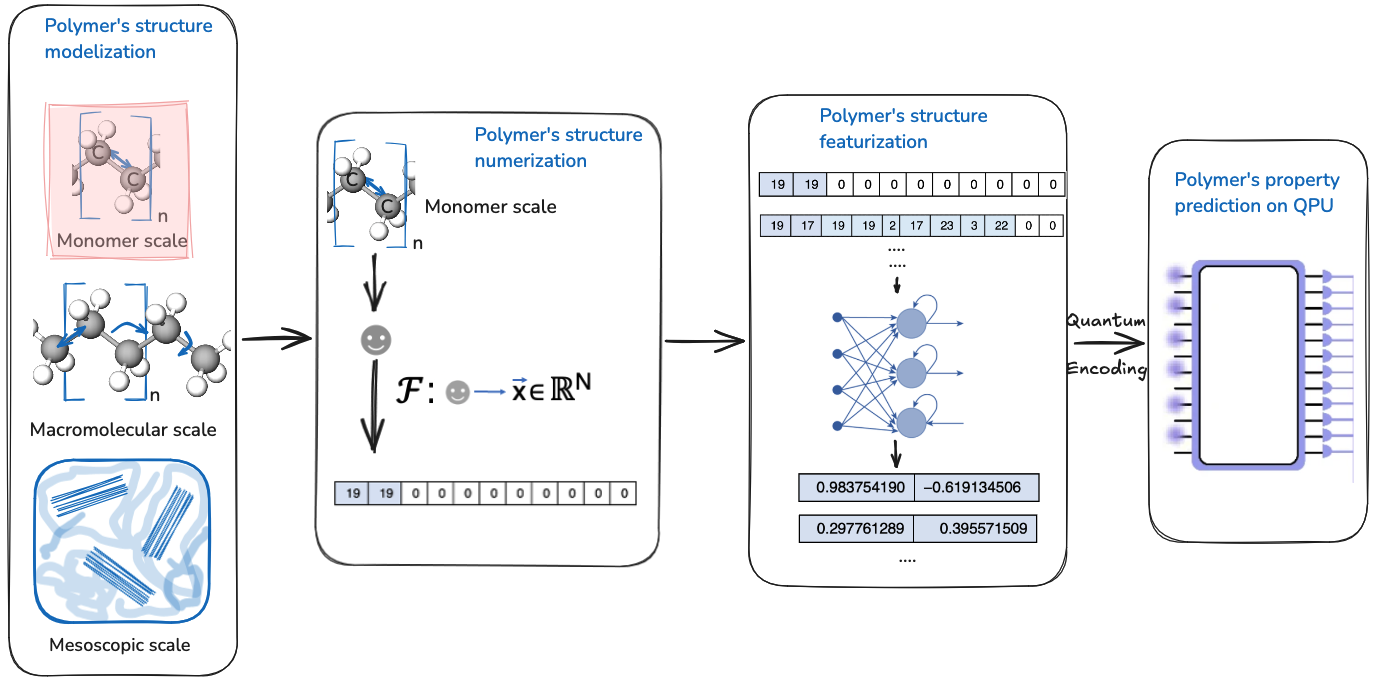}
    \caption{End-to-end schematic workflow for polymer property prediction using quantum machine learning.~Starting from multiscale polymer models (monomer, macromolecular, or mesoscopic scales), monomer structures represented by SMILES strings are converted into numerical vectors and transformed into $n$-dimensional feature vectors through a neural-network-based featurization step, followed by quantum encoding (embedding) as depicted in Fig.~\ref{fig:feature_pipeline}.The resulting quantum states are then fed into a quantum processing unit (QPU) or a noisy simulator to estimate polymer properties.}
\label{fig:general_pipeline}
\end{figure*}

\begin{figure*}[!htb]
    \centering
\includegraphics[width=1.0\textwidth]{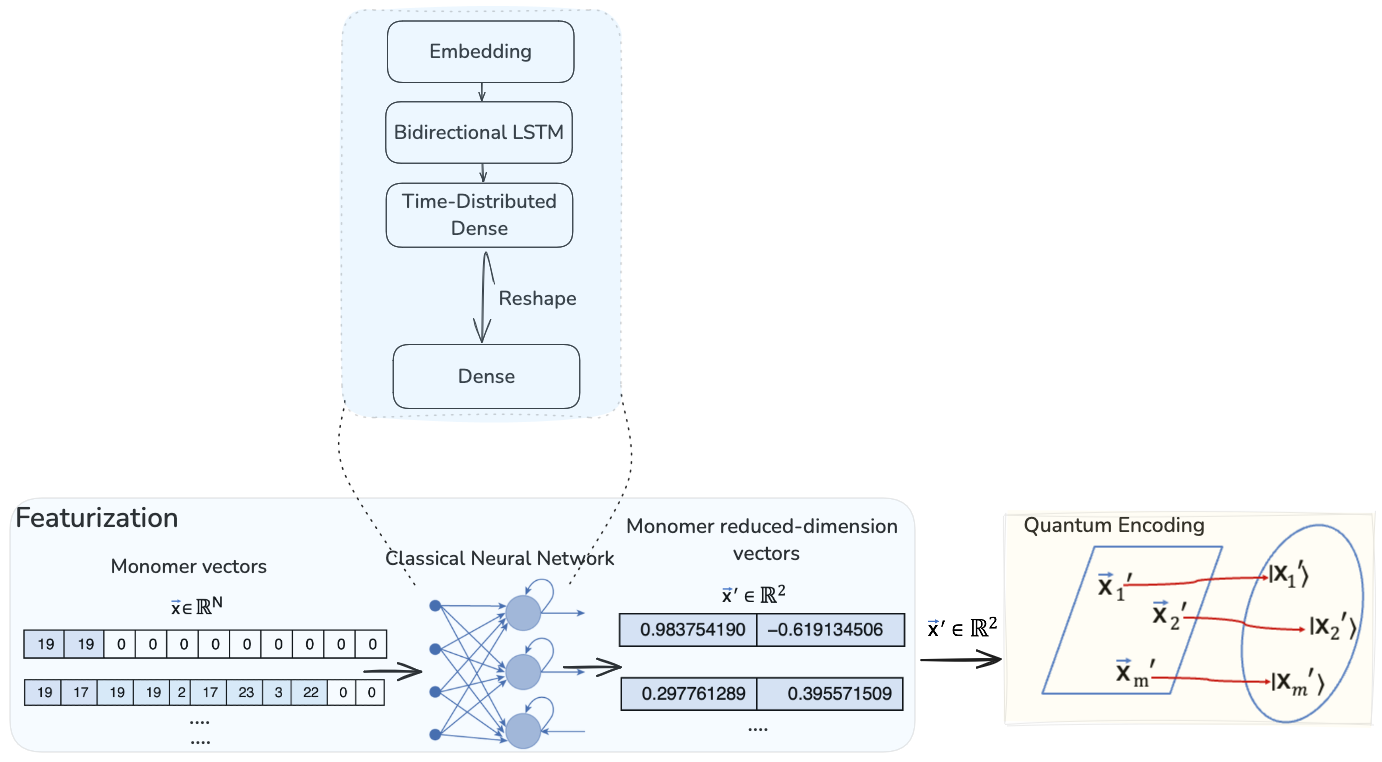}
    \caption{Overview of the polymer‐sequence featurization and encoding pipeline.~Monomer sequences are first converted into fixed-length numerical vectors.~A classical neural network-consisting of an Embedding layer, a Bidirectional LSTM, a Time-Distributed Dense layer, and a final Dense layer after reshaping-maps these monomer vectors $\bm{x}\in \mathbb{R}^N$ to low-dimensional latent representations $\bm{x}'\in \mathbb{R}^N$.~The resulting reduced-dimension monomer embeddings are then passed to the quantum pipeline, where each vector is encoded into corresponding quantum states, enabling hybrid classical-quantum downstream processing.}
\label{fig:feature_pipeline}
\end{figure*}

In this study, we employ such a photonic-native QML model to classify polymer species based on the size of their optical gaps.~The hybrid classification algorithm we employ integrates a single-photon-based quantum neural network \cite{Ascella_2023} with a classical deep neural network (DNN) feature extractor \cite{Polymer_paper}, which efficiently extracts relevant chemical information about polymer structures.~This extractor also reduces the dimensionality of the polymer vector representations, which are then embedded into a linear-optical quantum photonic circuit (QPC) using Fock-space encoding \cite{ref42}.~The output of the QPC is subsequently utilized for classification.~The end-to-end workflow of the method is shown in Fig.~\ref{fig:general_pipeline}, with a magnified view of the DNN-based featurization and quantum encoding in Fig.~\ref{fig:feature_pipeline}.

Our strategy defines a hybrid classical-quantum approach which allows for the use of simpler input Fock states with $n$-mode ($n$ - number of photons) linear-optical QPCs and a small number of photons.~The classical DNN feature extractor ensures in addition compact polymer representations containing ample chemistry information.~The approach was inspired by transfer learning and hybrid classical-quantum neural networks \cite{maria, TransferLearning, Quantum_Preskill}, widely exploited in the noisy intermediate scale quantum (NISQ) era.~In this work, we show that a hybrid classical-quantum photonic model can successfully capture structure–property correlation patterns in polymers, with its performance further supported by a proof-of-principle run on Quandela’s quantum processing unit (QPU).

The paper is organized as follows.~After a brief overview of the photonic-native QML formalism from Ref.~\cite{ref42}, with emphasis on the variational quantum classifier (VQC) approach in Sec.~\ref{sec:VQC}, we outline in Sec.~\ref{sec:experimental_setup} the setup used to implement this formalism on Quandela’s QPU Ascella~\cite{Ascella_2023}. Section~\ref{sec:polymers} then introduces the polymer-gap classification task, beginning with a concise description of the polymer data and encoding scheme (Sec.~\ref{sec:data_encoding}), followed by the construction of the classical DNN feature extractor in Sec.~\ref{sec:DNN_extractor_theory}.~The pipeline of the hybrid classical-quantum approach is described in Sec.~\ref{sec:HCQ}.~Results are next presented in Sec.~\ref{sec:results} before drawing conclusions in Sec.~\ref{sec:conclusions}. 

\section{Photon-based quantum neural network for classification tasks}\label{sec:QNN}
The photon-native NISQ variational classifier (Sec. \ref{sec:VQC}) developed in Reference \cite{ref42} is applied here to the classification of polymer gaps.~The formalism uses parameterized linear-optical QPCs with a specific architecture.~These QPCs are built of two trainable, parameterized beam splitter blocks $\mathcal{W}^{(1)}(\bm{\theta}_1)$ and $\mathcal{W}^{(2)}(\bm{\theta}_2)$ shown in Fig.~\ref{fig:QPU_architecture}, positioned at the input and output of the circuit, and a single, central data encoding block.~The classical data $\mathbf{x}$ is encoded using one phase shifter, encoding layer $S(\bm{x})$, per dimension of the data point\footnote{For example, for a $k$-dimensional data point $\mathbf{x}$, each component $x_l$ is encoded into the phase of a separate phase shifter}.~Reference~\cite{ref42} shows that for a fixed number of encoding phase shifters embedding classical data into a higher-dimensional Fock space using linear optics enhances the parametrized quantum circuit expressivity.~Because each phase shifter simultaneously uploads input data onto multiple Fock basis states, similar function-approximation performance can be achieved with fewer encoding layers, reducing circuit depth without requiring nonlinear components.~To carry out the classification task, we utilized the implementation of the VQC method on Quandela's QPU Ascella \cite{Ascella_2023}.~This implementation enables also to conduct quantum experiments on actual quantum hardware.~Its efficacy has already been demonstrated in Ref.~\cite{Ascella_2023} by successfully classifying the simple Iris dataset \cite{fisher1936iris}.~First, we recall briefly the VQC method in Sec. \ref{sec:VQC}.
\subsection{Variational Quantum Classifier}\label{sec:VQC} 

The $n$-photon QML model, produced by the parameterized linear-optical QPC, is defined as the expectation value of some observable measured using either photon-number resolving (PNR) or threshold detectors.~It is given by the Fourier series \cite{ref42}
\begin{eqnarray}
 f^{(n)}(\bm{x}, \bm{\Theta}, \bm{\lambda}) &=& \sum_{w \in \Omega_n}  c_w(\bm{\Theta}, \bm{\lambda})e^{iw.\bm{x}} 
\end{eqnarray}
with a frequency spectrum $ \Omega_n$ depending on the number $n$ of input photons and Fourier coefficients $\{c_w\}$ depending on
the trainable circuit block’s (beam splitter meshes) parameters
$\bm{\Theta} = (\bm{\theta}_1, \bm{\theta}_2)$ and observable parameters $\bm{\lambda}$.~In the following, we have considered that measurements were made with either PNR or single-photon (threshold) detectors.~We also set the number of photons $n$ to 3 and the number of modes $m$ to 5, which aligns with the setup used on Ascella (see Sec.~\ref{sec:experimental_setup}).~Then the $3$-photon QML model is defined as the expectation value of a trainable observable $\mathcal{M}(\bm{\lambda})$ with respect to the quantum state prepared by the QPC.~In this measurements setup $\mathcal{M} (\bm{\lambda})$ is diagonal in the Fock state basis and thus, the model reads
\begin{equation}
f^{(3)}(\bm{x}, \bm{\Theta}, \bm{\lambda}) =
\sum_{{n_i}} \lambda_{{n_i}} |\langle {n_i}| \mathcal{U}(\bm{\Theta}, \bm{x})|{n^{in}}\rangle|^2,
\end{equation} where \begin{equation}
\mathcal{U}(\bm{\Theta}, \bm{x}) = \mathcal{W}^{(2)}(\bm{\theta}_2)\mathcal{S}(\bm{x})\mathcal{W}^{(1)}(\bm{\theta}_1).
\end{equation} Here, $|{n^{in}}\rangle = |n_1, n_2, n_3, n_4, n_5\rangle$ is the input 3-photon Fock state with $\sum_{j=1}^{m=5} n_j = 3$ ($n_j$ is the number of photons in mode $j$). The trainable observable equals \begin{equation}
\mathcal{M}(\bm{\lambda}) = \sum_{{n_i}} \lambda_{{n_i}} |{n_i}\rangle \langle{n_i}|,
\end{equation} where each output state $n_i = |n_k,n_l,n_o,n_p,n_q\rangle$ obeys the equality $n_k + n_l + n_o + n_p + n_q = 3$.~The unitary transformation $\mathcal{U}(\bm{\Theta}, \bm{x})$ is parameterized by $\bm{\Theta}$ and it is a product of the single data-encoding layer $\mathcal{S}$ and the two trainable beam splitter meshes $\mathcal{W}^{(1, 2)}$.

The model is trained by minimizing variationally the regularized squared loss function over $\bm{\Theta}$ and $\bm{\lambda}$ \cite{ref42}
\begin{equation}\label{loss_VQA}
    \mathcal{L}(\bm{\Theta}, \bm{\lambda}) = \frac{1}{2 N} \ \sum_{l = 1}^{N} (y_l - f^{(3)}(\bm{x}_l, \bm{\Theta}, \bm{\lambda}))^2 + \alpha \bm{\lambda} \cdot \bm{\lambda}, 
\end{equation}
where $\{y_l\}_{l=1}^{N}$ are the true, measured data point labels, $N$ is the size of the train dataset $\{\bm{x}_l, y_l\}_{l=1}^{N}$ and $f^{(3)}(\bm{x}_l, \bm{\Theta}, \bm{\lambda})$ is the $3$-photon QML model.~The term $\bm{\lambda} \cdot \bm{\lambda} = \sum_l \lambda_l^2$, the sum of squared observable parameters, is a regularization term with a weight $\alpha$.~In practice, we use a seesaw optimization approach suited to the experimental implementation of the VQC that also permits to handle the strong non-convexity of the loss function, see Sec.~\ref{sec:comp_details}.~The predicted class of the polymer data points is then determined by the trained model's sign given by the circuit output
\begin{equation*}
f^{(3)}_{sign} (\bm{x})  =  sgn[f^{(3)}(\bm{x}, \bm{\Theta}_{opt}, \bm{\lambda}_{opt})],
\end{equation*}
with the optimized parameters for the circuit and observable $\bm{\Theta}_{opt}$ and $\bm{\lambda}_{opt}$, respectively.
% Remind of what variational approach is.

\subsection{Setup for a photon native computation}
\label{sec:experimental_setup}
    \begin{figure*}[!ht]
        \centering \includegraphics[scale=0.30]{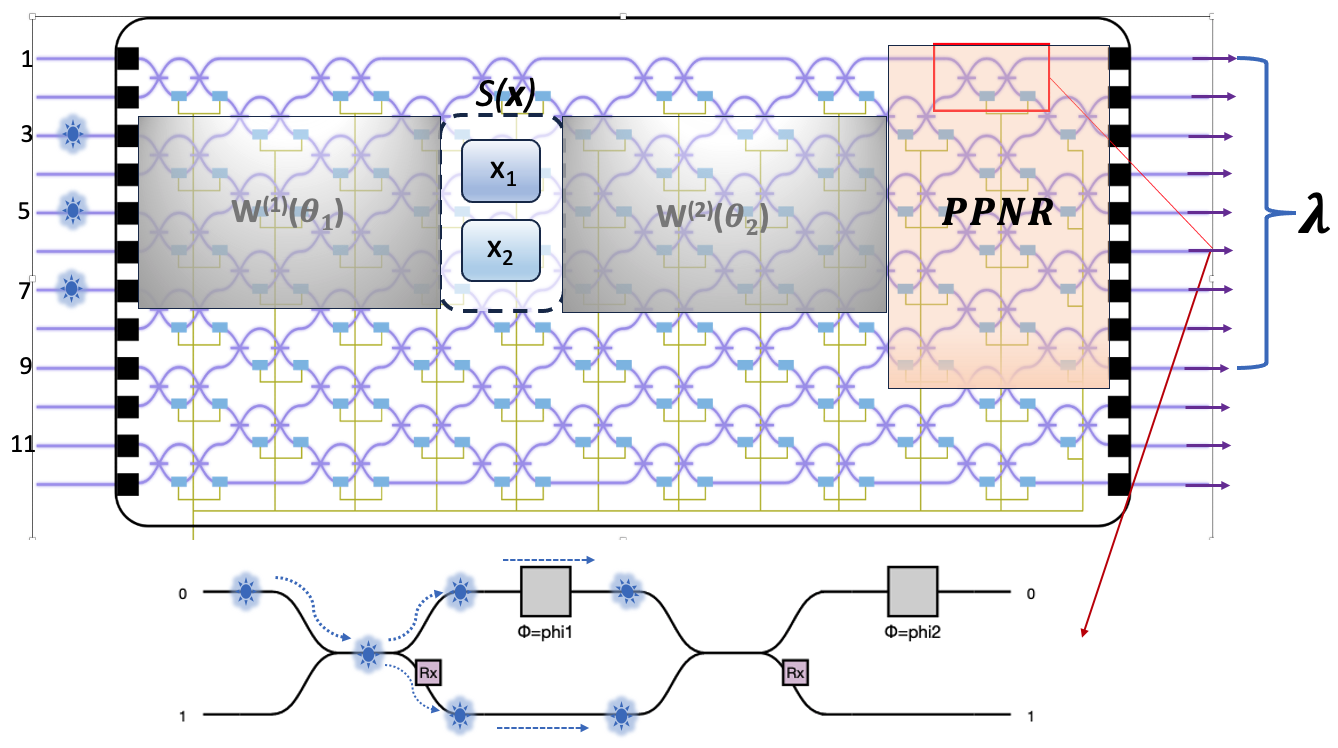}
\caption{\textit{Reproduced from Fig.~S11 in Ref.~\cite{Ascella_2023}, with modifications}: Linear optical QPC implemented on Qunadela’s QPU Ascella for the VQC approach.~Spatial modes 3 to 7 are selected out of the 12 available modes to form the first parametrizable (by $\bm{\theta_1}$) block on the QPC by exploiting 16 of
the re-configurable phase shifters.~The second parametrized (by $\bm{\theta_2}$) trainable block with 16 phase shifters is followed by a pseudo PNR (PPNR) detectors
block (in orange) that spreads over four extra spatial modes.~By design, four phase shifters acting on spatial modes 4 to 7 are attributed to encode four-feature data block, sandwiched between the two trainable blocks (in grey).~The phases of the chip that have neither trainable nor data encoding functions are set to 0.
}
    \label{fig:QPU_architecture}
        \end{figure*}
For the hardware-based test, we employed the general-purpose single-photon-based quantum computing platform Ascella described in Ref.~\cite{Ascella_2023}.~Ascella is a versatile quantum computing machine that supports quantum neural network (QNN) training.~It features a quantum dot-based single-photon source connected to a reconfigurable 12-mode integrated photonic chip.~Its architecture, shown in Fig.~\ref{fig:QPU_architecture}, is reproduced from Fig.~S11 of Ref.~\cite{Ascella_2023}.~Figure modifications highlight the QPU spatial modes selected to implement the VQC algorithm.

The circuit on the chip consists of a universal interferometer whose output is measured with threshold single-photon detectors.~For performing  supervised-learning classifications, the VQC algorithm \cite{ref42} described in Sec.~\ref{sec:VQC} was implemented on the photonic quantum machine  \cite{Ascella_2023} in order to construct the QNN using a native photonic ansatz.~The latter is designed by using beam splitters and phase shifters which specific topology is shown in Fig.~\ref{fig:QPU_architecture}.~The ansatz follows the architecture of the linear-optical QPC proposed in Ref.~\cite{ref42} with two parameterizable (by $\bm{\theta}_1$ et $\bm{\theta}_2$) blocks and a data-encoding block sandwiched in between.~A detailed description of the dedicated Ascella's chip topology is given in the original work~\cite{Ascella_2023}.~The platform uses a pseudo–photon-number-resolving (PPNR) block to achieve photon number resolution, where several single-photon detectors distributed over multiple modes operate in parallel.~Photon counts observed across several modes are aggregated and interpreted as a collective photon count for a single effective mode.

In the numerical CPU-based simulations and the hardware-based test, an initial 3-photon input Fock state $|001010100000\rangle$ is prepared defined over the 12 modes of the architecture in Fig.~\ref{fig:QPU_architecture}.~The 3 single photons enter at spatial modes 3, 5, and 7, respectively.~The setup of phase shifters and beam splitters prepare different output 
quantum states defined by the photon counts measured at each detector at the end of the workflow.~The $\bm{\lambda}$ parameters of the observable $\mathcal{M}$ are assigned as weights to each observed output quantum state.
Similar to classical NN, the parameters of the QNN $\bm{\theta}$, but also the weights $\bm{\lambda}$, are optimized here classically to minimize the loss function in Eq.~(\ref{loss_VQA}).~Within a conventional VQC optimization cycle, the workflow prepares a quantum state that depends on a set of parameters ($\bm{\theta}, \bm{\lambda}$) which are next optimized within a classical setup.~The optimization strategy is described in Sec.~\ref{sec:comp_details}. 
\section{Polymer gap classification}\label{sec:polymers}
In this section, we present the polymer classification task within the context of the hybrid classical-quantum approach introduced in Ref. \cite{Polymer_paper}.

\subsection{Polymer data and encoding}\label{sec:data_encoding}
In the present work, we employ the synthetic dataset originally developed for our study on photonic quantum approaches to polymer classification~\cite{Polymer_paper}.~High-quality experimental databases for polymers are scarce and often sparse as measuring polymer properties is a resource and time consuming task~\cite{data_driven, ref31, ref32}.~To ensure an optimal amount of data for our approach, we utilise data yield by 
Density Functional Theory (DFT) calculations \cite{ref4, ref37} on monomers or oligomers employing either B3LYP \cite{ref38, ref39} or CAM-B3LYP \cite{ref40} functionals.~The dataset is derived from a database containing optoelectronic properties of oligomers for organic photovoltaic (OPV) applications, as well as electronic properties extrapolated to the polymer limit~\cite{ref37}.~Here, we focus on the subset of extrapolated polymer gap $E_{g}$~\footnote{For a monomer, the gap is equal to its first excitation energy that, at first approximation, equals $\varepsilon_{LUMO}$ - $\varepsilon_{HOMO}$.}.~After pre-processing (see Appendix~\ref{sec:polymer_data}), we obtained a dataset of 52815 polymers with their extrapolated gaps.~The polymer band gaps were then grouped into two classes-visible (VIS) and near-infrared (NIR)-according to their magnitude.~Our objective is to show that the hybrid classical–quantum model-trained and evaluated on the Ascella platform-can accurately classify similar, but unseen polymer species.

To represent the polymer structure in a machine readable form, each species is first described by a Simplified Molecular Input Line Entry Specification (SMILES) 
of its monomer (or oligomer).~Utilizing the monomer is deemed sufficient for ground-state properties such as $E_{g}$.~SMILES is a compact chemical line notation that represents a chemical structure in a computer-adapted ASCII strings form~\cite{Daylight_SMILES_Theory}.~Using a vocabulary of atom, bond, ring and stereometry symbols (syntax rules) with chemically specific semantic rules, we can, for example, build the SMILES of polystyrene's monomer as C=CC1=CC=CC=C1.~Here, "C" denotes a carbon atom and "=" signifies a double bond between two C atoms.~The number "1" indicates the C atoms involved in closing the ring in the compound.~Hydrogen atoms are usually omitted, but are present implicitly according to atoms standard valences.

Each SMILES string undergoes further transformations in order to cast it into a four-feature (4-component) input vector for the photon-native VQC.~To derive those polymer 4-component numerical representations, we have adopted 
the chemical
language processing model by Chen \textit{et al.}~\cite{ref5}.~Because SMILES is a chemical linguistic form that enables expressing a chemical structure as a text, the model borrows text encoding strategies and DNN architectures from the
Natural Language Processing domain.~More specifically, the monomer SMILES is viewed as a sequential string of characters, and a Recurrent Neural Networks (RNN) \cite{ref43} are thus employed as being adapted to such input data. 

Before being processed by the DNN, that is here of RNN type, each SMILES is cast into an integer-value vector according to the procedure described in appendix \ref{sec:Encode_smiles}.~This involves tokenization of the SMILES, construction of unique tokens dictionary, and encoding of each SMILES tokens depending on their positions in this dictionary.~The resulting integer-valued vectors, each of length 139,\footnote{Here, 139 corresponds to the maximum monomer length; shorter monomers are padded with zeros to reach a length of 139.} provide a simple feature representation of the chemical species but do not necessarily capture all relevant structural information.~We have thus combined them
with a subsequent character embedding, as proposed in Refs.~\cite{ref5, Polymer_paper}.~The embedding addresses the otherwise poor accuracy of DNN models associated with the use of simple integer-value vectors \cite{ref44}.~In practice, this embedding is  realized by the embedding layer of the classical RNN-based feature extractor employed in this study.~Its architecture is depicted in Fig.~\ref{fig:lstm}.~We describe it in some details in the following section \ref{sec:DNN_extractor_theory}. 

\subsection{Classical DNN feature extractor}\label{sec:DNN_extractor_theory}
We now describe the architecture of the classical DNN feature extractor which workflow is shown in Fig.~\ref{fig:feature_pipeline}.~The extractor is built from recurrent layers of the bidirectional Long Short-Term Memory (LSTM) type~\cite{ref43} (see Fig.~\ref{fig:lstm}).~Given the sequential form of the chemical data, LSTMs are well suited to extract the relevant structure-property patterns.

To generate the DNN feature extractor, the model was first trained on a subset of 52815 input vectors, each comprising 139 components.~We randomly sampled 90\% of this dataset, corresponding to 47534 polymer structures, ensuring that the subset was balanced across the VIS and NIR polymer classes.~This balanced subset was then split into training, validation, and test sets in an 80:10:10 ratio, a common practice in machine learning.~The DNN was subsequently trained\footnote{Training the DNN involves adjusting the network’s weights and biases to minimize a predefined loss (or cost) function that quantifies the difference between the predicted and true classes for each input sample.} on the training subset for a binary classification task, while monitoring the decrease of the training loss function over time.~This loss function measures the discrepancy between the model’s predicted classes and the actual labels in the training data. The model was trained for the maximum of 100 epochs with an Early Stopping mechanism that, at each epoch, evaluated the DNN’s accuracy on the validation subset to prevent overfitting.~The Early Stopping uses the validation loss function~\footnote{The validation loss function is analogous to the train loss, but it is defined over the validation subset.} as a metric to halt the training if the loss does not decrease during a pre-defined number of epochs, called  \textit{patience}.~Here the patience is set to 15.~This prevents the trained model from overfitting on the validation and test subsets.~The optimizer used to update the network weights at each epoch is a stochastic gradient descent method, called Adam optimizer.~The 
learning rate, that is the tuning parameter in the optimization algorithm, had the default value of 0.001.~Within these settings, the DNN was trained effectively for 56 epochs.~The resulting trained model achieves an accuracy of 0.879 on the test set, computed as the ratio between the number of correctly predicted classes and the total number of predictions.

The classical DNN feature extractor is obtained by removing the last three dense layers shown in Fig.~\ref{fig:lstm} from a trained DNN model.~In this set of calculations, we first consider a DNN feature extractor whose first dense layer outputs a two-dimensional representation (\(k = 2\); see Fig.~\ref{fig:lstm}).~This choice was originally dictated by an earlier implementation of the VQC method.~To interface with the photonic VQC implementation, which requires four dimensional input data vectors (see, Fig.~\ref{fig:QPU_architecture}), this representation was augmented to \(k = 4\) via the mapping \((x_1, x_2) \mapsto (x_1, x_2, x_1^2, x_2^2)\).~In addition, two independent DNN extractors producing four-dimensional outputs (\(k = 4\) in Fig.~\ref{fig:lstm})) were trained directly.~These additional models serve to assess the robustness of the results and to verify that the augmented representation does not introduce bias.~The training was done for 48 and 53 epochs using Early Stopping and the accuracy of the trained models on the test set was 0.869 and 0.850, respectively.      

Applied to the remaining 10\% of the dataset (a balanced subset (representative of the two polymer classes) of 5281 polymers out of 52815 input vectors, each with 139 components), the DNN feature extractor captures rich chemical information while reducing the input dimensionality to $k$-component vectors.~In fact, the embedding layer of the classical DNN in Fig.~\ref{fig:lstm} first transforms the polymer data into compact, dense representations, which are then further processed and dimensionally reduced by the subsequent neural network layers.~Further details on the DNN feature extractor are provided in Appendix~\ref{sec:DNN_extractor}.      
\subsection{Hybrid classical-quantum approach}\label{sec:HCQ}
In this section, we assemble the classic DNN feature extractor of Sec.~\ref{sec:DNN_extractor_theory} with the photon-native VQC of section \ref{sec:experimental_setup} to obtain a hybrid classical-quantum approach.~The approach was already discussed in our previous work~\cite{Polymer_paper}. Here, we briefly summarize it for completeness.~In our studies, we used the implementation from Ref.~\cite{Ascella_2023}, developed with Perceval, the open-source framework by Quandela for programming photonic quantum computers~\cite{ref36}.
\begin{figure}[!htb]
    \centering
\includegraphics[width=1.0\linewidth]{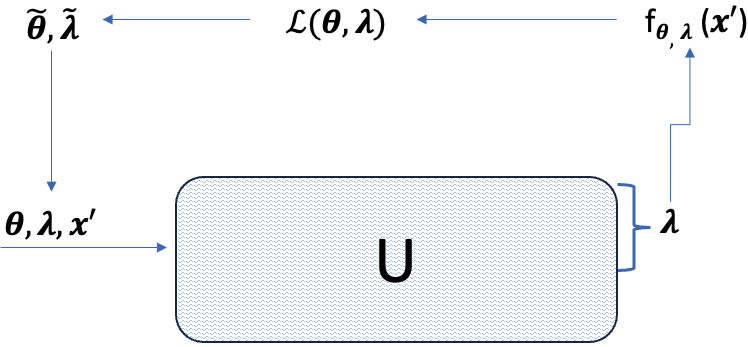}
    \caption{Schematic of the learning loop for the photonic variational quantum classifier (\textit{modified version of Fig.~S11 in Ref.~\cite{Ascella_2023}}).~A photonic circuit specified by the unitary $U$ is parametrized by ($\bm{\theta}$, $\bm{\lambda}$)
fed with the encoded input $\bm{x'}$.~The measurement statistics provide the model prediction (output) \( f_{\bm{\theta},\bm{\lambda}}(\bm{x'}) \), from which the loss \( \mathcal{L}(\bm{\theta},\bm{\lambda}) \) is computed and used to update the parameters to
($\bm{\tilde\theta}$, $\bm{\tilde\lambda}$) for the next optimization step.~$\bm{\lambda}$ are the variational parameters of the model.
}
    \label{fig:VQC}
\end{figure}

The role of the classical DNN feature extractor is two fold.~It extracts pertinent 
chemical information from the polymer data and then it compacts it into dense feature vector representations $\bm{x}'$.~It can thus be viewed as an input data-encoding and dimension reduction technique (see, Fig~\ref{fig:feature_pipeline}).~In analogy with other hybrid classical-quantum neural networks~\cite{TransferLearning}, it transfers information from the classical to the quantum part of the network.~The extracted feature vectors $\bm{x}'$ serve as input data for the VQC algorithm which learning pipeline is depicted in Fig.~\ref{fig:VQC}.~As illustrated in Fig.~\ref{fig:feature_pipeline}, the compound features, or latent representations \(\bm{x}'\), are initially transformed through a quantum encoding step that maps them to quantum states in this pipeline.~Notably, the modular nature of this hybrid approach enables the integration of the DNN feature extractor with various quantum classification methods, including quantum kernel techniques.~Some details on the optimization of the parameters ($\bm{\theta}$, $\bm{\lambda}$) of the VQC are provided in Sec.~\ref{sec:comp_details}. 

\section{Computational details}\label{sec:comp_details}
In this study, we performed binary polymer classification tasks using two types of datasets.~The first type consists of data from 5281 polymers encoded by the DNN feature extractor with $k$ = 2 or $k$ = 4.~The DNN feature extractor with $k$ = 4 produces inherently 4-dimensional vectors 
$\bm{x'} = (x'_1, x'_2, x'_3, x'_4)$.~The data feature vectors obtained with $k$ = 2 are depicted in Fig.~\ref{fig:2D_input_data_big_data_set}.~The figure shows the distribution of the two-dimensional feature vectors.~These distributions highlight how the two classes occupy different regions of the feature space while still exhibiting overlap along individual feature vectors' dimensions (coordinates) as seen from the overlap between the Gaussian marginals\footnote{This observation highlights that neither feature coordinate alone provides complete class separability.}.~The figure illustrates the geometry of the learned feature space, showing that effective classification relies on the joint distribution of feature coordinates rather than on individual marginal statistics.

As explained in Sec.~\ref{sec:DNN_extractor_theory}, we have augmented the two-dimensional (two-component) vectors to four-dimensional, 4D (four-component) counterparts in order to adapt their structure to the experimental implementation of the photon-native VQC.~This strategy also permits to evaluate the impact of the compression of the initial data.~Alternative augmentation strategies are possible, but for the sake of this demonstration we retained a simple mapping \((x'_1, x'_2) \mapsto (x'_1, x'_2, x_1'^{2}, x_2'^{2})\).~The data normalization before the augmentation  by subtracting mean and dividing by standard deviation in this specific case did not affect the training outcome.~Thus we work with normalized vectors.~Each vector component is then encoded using a single phase shifter.

The second type of dataset comprises 557 polymers represented by $k$-dimensional feature vectors (for
$k$ = 2
 see Fig.~\ref{fig:2D_input_data}).~These feature vectors are randomly sampled from the larger datasets described above.~The resulting reduced datasets remain representative of the original distributions and are balanced with respect to the two polymer classes, as illustrated in Fig.~\ref{fig:Distributions_small_data_set} for the case 
$k$ = 2.~Using the same augmentation strategy described previously, an augmented four-dimensional dataset is also constructed.~Each 4D dataset is then split into training and test subsets using a 75:25 ratio.~Prior to splitting, the data are shuffled and stratified to preserve the class balance in both subsets.~Note that the 4D feature vectors are normalized by construction.

The role of data augmentation here is similar to its use in classical machine learning, where embedding data samples into a higher-dimensional feature space can facilitate class separation and improve classification performance~\footnote{By increasing the dimensionality of the feature representation, the augmented data allow the classifier to capture more expressive decision boundaries.}.~Due to limited computational budget, we validate on the QPU the datasets with augmented features only.

The two classes VIS and NIR are further encoded as labels 1 and -1, respectively using a categorical encoding.

For the optimization of the trainable $\bm{\theta}$ and observable $\bm{\lambda}$ parameters, a cyclic see-saw optimization strategy was adopted as in Ref.~\cite{Ascella_2023}.~The trainable parameters $\bm{\theta}$ and the observable parameters $\bm{\lambda}$ are optimized into two separate alternating loops with the objective to find optimal $\bm{\lambda}$-s for each fixed set of $\bm{\theta}$ values.~This separation is motivated by the different computational costs: adjusting $\bm{\theta}$ requires reconfiguring the photonic circuit, whereas optimizing $\bm{\lambda}$ requires only classical post-processing~\cite{Ascella_2023}.

The optimization over the trainable parameters $\bm{\theta}$ is performed using Gaussian process-based Bayesian optimization.~The Gaussian process acts as a probabilistic surrogate model for the cost function over $\bm{\theta}$ enabling sample-efficient optimization under experimental constraints\footnote{The Gaussian process is not modelling the data, or the classifier, it is modelling the objective function landscape over 
$\bm{\theta}$.}.~For each proposed value of $\bm{\theta}$, the observable weights $\bm{\lambda}$ are optimized using the Nelder–Mead method.

Any computational trial has been run for 15 iterations.~In the proof-of-principle run on Quandela's QPU Ascella, each iteration required 10 QPU executions, corresponding to one execution per training-data batch (10 batches are used) with 50000 samples/shots per execution.~Because the loss function exhibits a  nonconvex, multimodal landscape, different optimization runs may converge to distinct local minima.~To account for this variability, each computational trial was repeated five times, and the reported accuracies correspond to averages over these independent runs.~Both the  simulations on the classical machine and the experiments on the quantum device were performed using the photon-native implementation of the variational quantum classifier developed in Ref.~\cite{Ascella_2023}, built on the Perceval framework. 

\section{Results}\label{sec:results} 
     \begin{figure*}[!htp]
         \centering
         \begin{subfigure}[b]{0.4\textwidth}
             \centering
             \includegraphics[width=\textwidth]{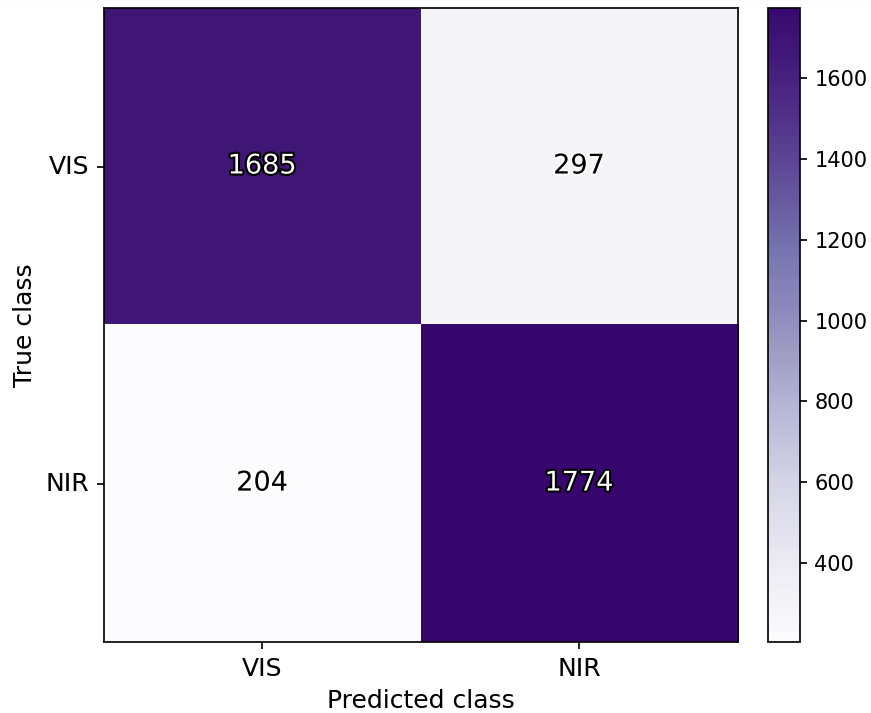}
             \caption[Train dataset of 3960 species.]%
             {{\small Train dataset: 3960 (augmented) vectors.}}    
             \label{fig:}
         \end{subfigure}
         \hfill
         \begin{subfigure}[b]{0.4\textwidth}   
             \centering 
 \includegraphics[width=\textwidth]{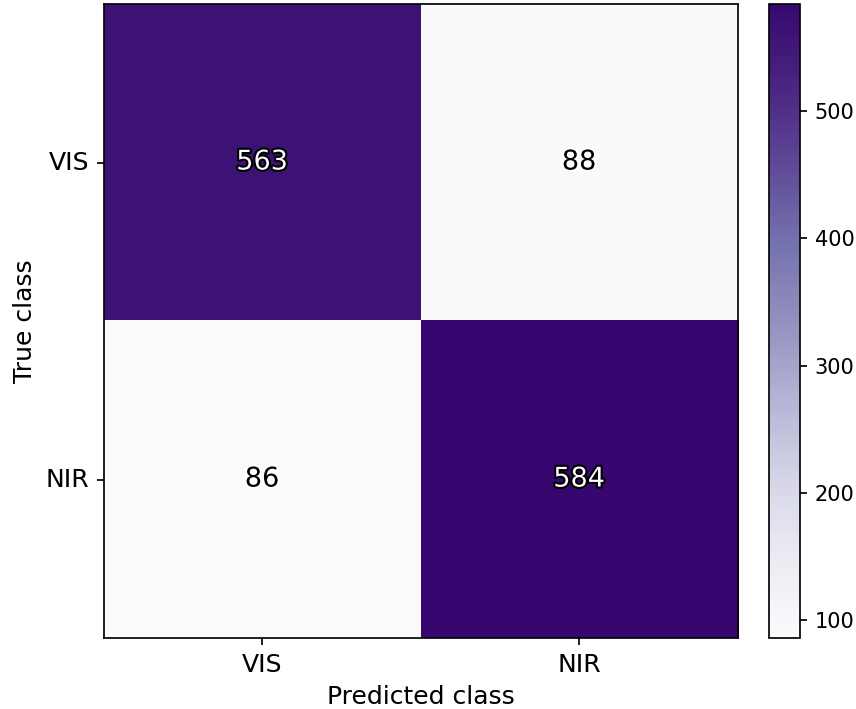}
             \caption[Test dataset of 1321 species]%
             {{\small Test dataset: 1321 (augmented) vectors.}} 
             \label{fig:}
         \end{subfigure}
         \hfill
         \begin{subfigure}[b]{0.4\textwidth}   
             \centering 
 \includegraphics[width=\textwidth]{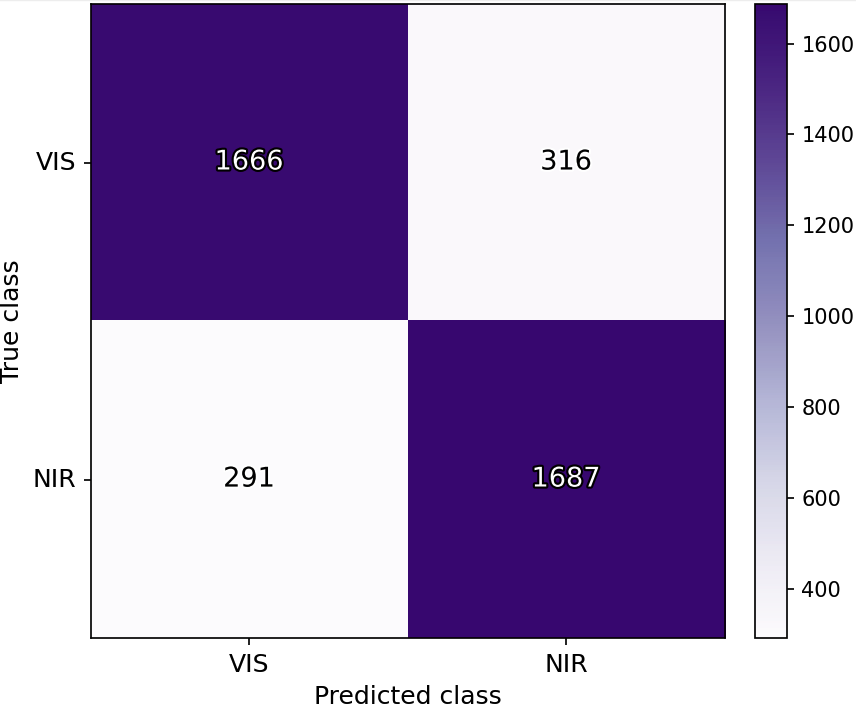}
             \caption[Test dataset of 3960 species]%
             {{\small Train dataset: 3960 vectors.}} 
             \label{fig:}
         \end{subfigure} 
         \hfill
                 \begin{subfigure}[b]{0.4\textwidth}   
             \centering 
 \includegraphics[width=\textwidth]{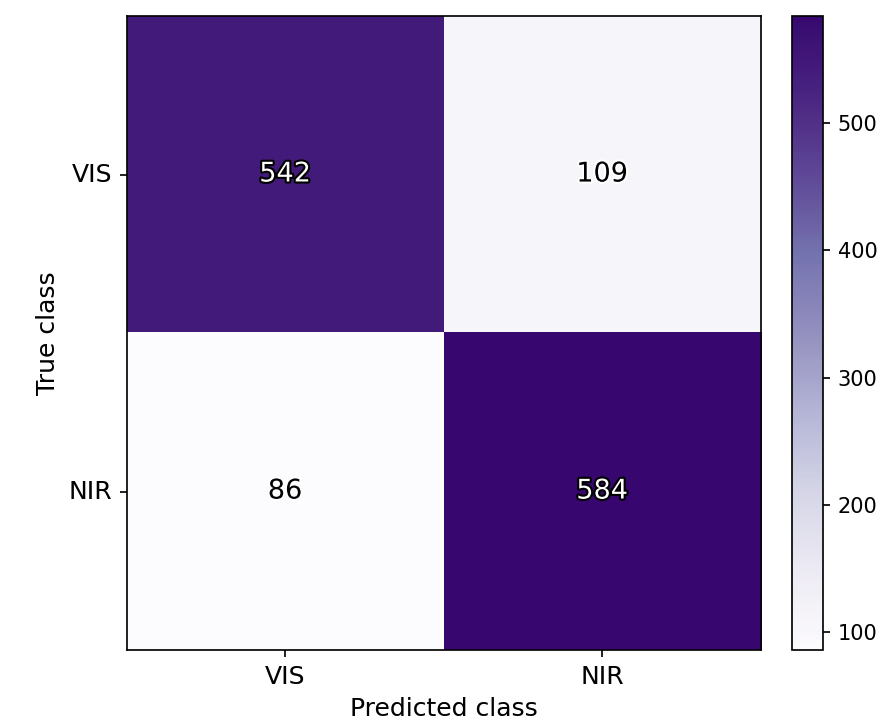}
             \caption[Test dataset of 1321 species]%
             {{\small Test dataset: 1321 vectors.}} 
             \label{fig:}
         \end{subfigure} 
         \caption[Confusion matrices for the polymer binary classification task on the dataset of 5281 species.]    
         {Confusion matrices for the binary classification of polymers in the dataset comprising 5281 species.~Each polymer is represented by a 4D feature vector encoding its monomeric chemical structure.~The feature vectors are obtained using the DNN-based feature extractor, either directly with $k$ = 4 or by $k$ = 2 features followed by data augmentation to construct a 4D representation.} 
 \label{fig:CM_results_large_data_set}
     \end{figure*}
In this section, we report on the performance of the VQC model from both simulations and proof-of-principle experiments.~As detailed previously, both the quantum circuit parameters and the observable coefficients are trained explicitly to optimize classification accuracy.

We first performed numerical simulations on a classical CPU for the binary classification of polymer datasets containing either 5281 or 557 samples.~Each polymer was represented by a four-dimensional ($k$ = 4) feature vector encoding its monomeric chemical structure, as described in Sec.~\ref{sec:comp_details}.\begin{table}[!ht]
    \caption{Mean training and test accuracies accuracies (acc.) with corresponding standard deviations (std.) obtained from CPU-based simulations of polymer binary classification.~Results are reported for datasets containing either 5281 or 557 polymer samples, represented by four-dimensional feature vectors.~The column labelled "Augment" indicates whether data augmentation was applied to obtain the 4D vectors.}
    \centering
    % \begin{tabular}{>{\hsize=0.6cm}p{0.6cm}|>{\hsize=1.3cm}p{1.3cm}|p{0.9cm}|p{0.9cm}|p{0.9cm}|p{0.9cm}}
    %\begin{tabular}{|c|p{1.3cm}|c|c|c|c|}
    \begin{tabular}{l c *{4}{c}}
    % Fails:
    % \begin{tabular}{p{0.6cm}|p{1.3cm}|p{0.9cm}|p{0.9cm}|p{0.9cm}|p{0.9cm}}
    \toprule
Data  &  Augment &\multicolumn{2}{c}{Train}     & \multicolumn{2}{c}{Test} \\
        & & acc.  & std.& acc.  & std.          \\
    \midrule
\hline 
       5281 &$\times$  &0.848 &0.004 & 0.843&0.006 \\
                5281 & $\checkmark$  &0.871 &0.002&  0.868&0.004 \\
        557 &$\times$ & 0.858& 0.012& 0.856&0.029 \\
                557 &$\checkmark$  &0.869 &0.011& 0.827&0.006 \\ \hline 
    \end{tabular}
    \label{tab:results}
\end{table}
Simulation results were obtained using the Strong Linear Optical Simulation (SLOS) backend \cite{Heurtel2023} of Perceval with a noisy source to simulate a real device experiment.~SLOS is a backend algorithm for simulating linear optical quantum computing processes.~It permits to simulate the output distribution of linear optical quantum circuits.~The optical losses of the single-photon source was set to 0.92, with 0.0 corresponding to a perfect, noiseless simulation.~The indistinguishability  parameter had a value of 0.92, where the maximal value of 1.0 would mean a pair of completely indistinguishable photons (perfect case).~All other source parameters were set to their default values.~Note, that the noise values were chosen to mimic the actual conditions on Ascella.

The resulting mean training and test accuracies and standard deviations obtained from the CPU simulations are summarized in Table~\ref{tab:results}.~We have also reported the confusion matrices in
Figs.~\ref{fig:CM_results_large_data_set} for all pairs (Data, Augment) in Table~\ref{tab:results}.~The pair (557, $\times$) is characterized by a large standard deviation for the test accuracy (see, third raw in Table~\ref{tab:results}) and hence, it is not straightforward to select a single computational trial out of the 5 experiments to report on.~For the other pairs, we have selected confusion matrices corresponding to single computational runs with the best (train,  test) pair accuracies.   

Upon examining the CPU simulation results for the dataset comprising 5281 polymers, we observe that our hybrid model achieves a mean accuracy comparable to that of the classical  DNN models employed in constructing the feature extractors.~The model
    \begin{figure*}[!htpb]
        \centering
        \begin{subfigure}[b]{0.4\textwidth}
            \centering
\includegraphics[width=\textwidth]{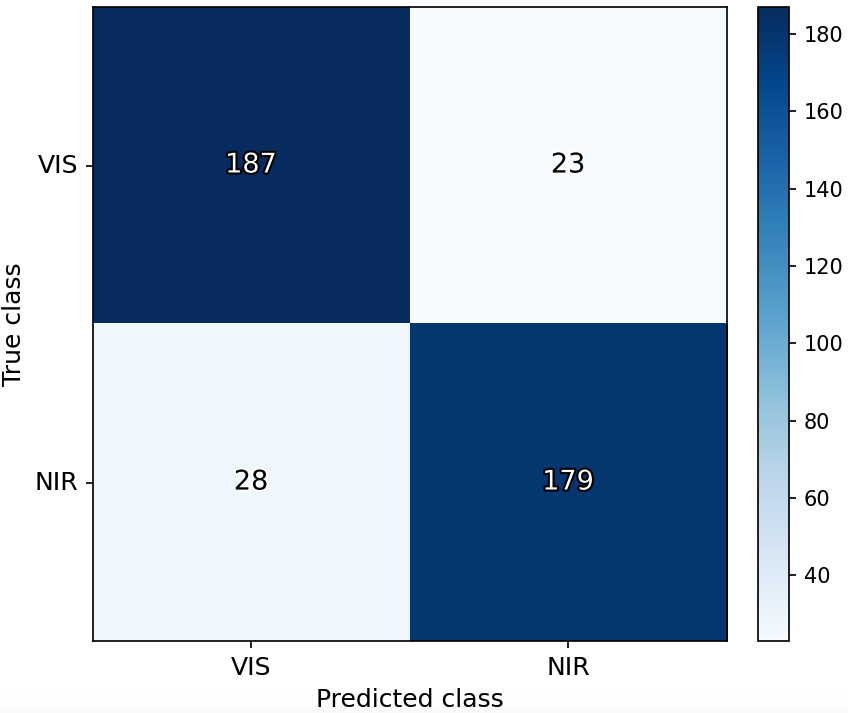}
            \caption[Train dataset of 417 species]%
            {{\small Train dataset: 417 4D (augmented) vectors}}    
            \label{fig:}
        \end{subfigure}
        \hfill
        \begin{subfigure}[b]{0.4\textwidth}   
            \centering 
            \includegraphics[width=\textwidth]{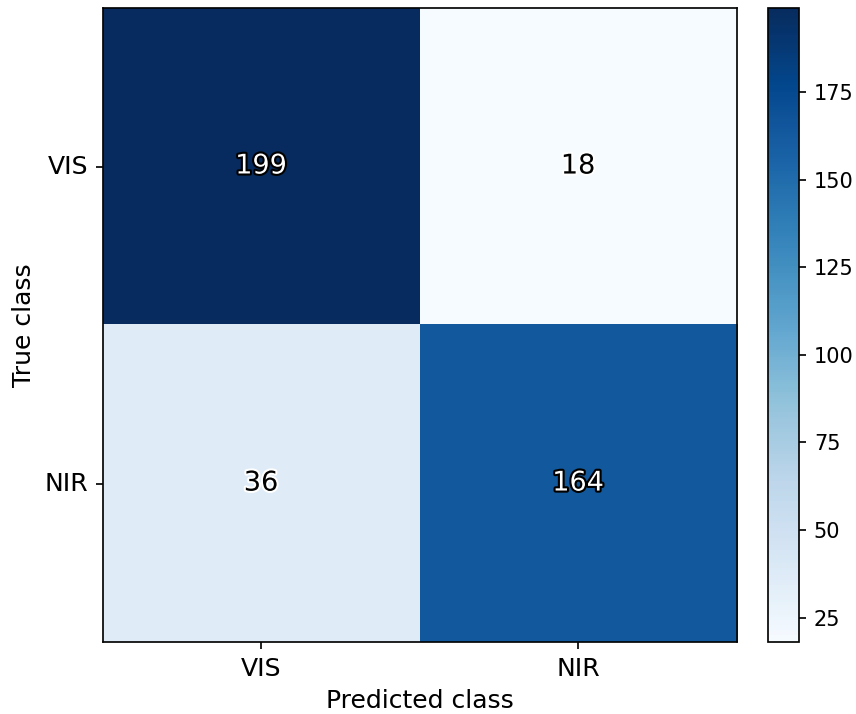}
            \caption[Train dataset: 417 4D (augmented) vectors]%
            {{\small Train dataset: 417 4D (augmented) vectors }} 
            \label{fig:}
        \end{subfigure}
        \hfill
        \vskip\baselineskip
        \begin{subfigure}[b]{0.4\textwidth}   
            \centering 
\includegraphics[width=\textwidth]{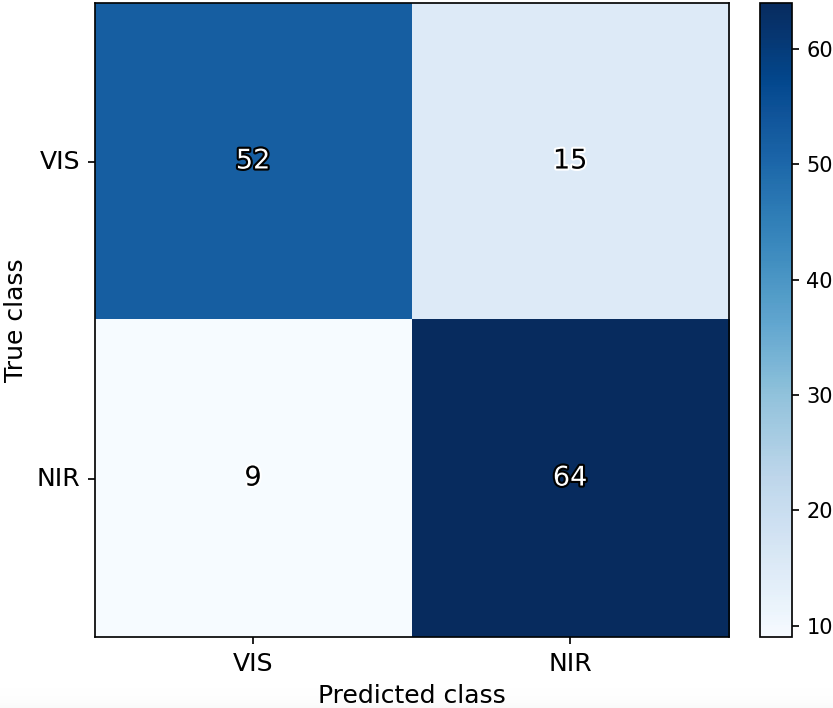}
            \caption[Test dataset: 140 4D-augmented vectors]%
            {{\small Test dataset: 140 4D (augmented) vectors}} 
            \label{fig:}
        \end{subfigure}
                \hfill
        \begin{subfigure}[b]{0.4\textwidth}   
            \centering 
            \includegraphics[width=\textwidth]{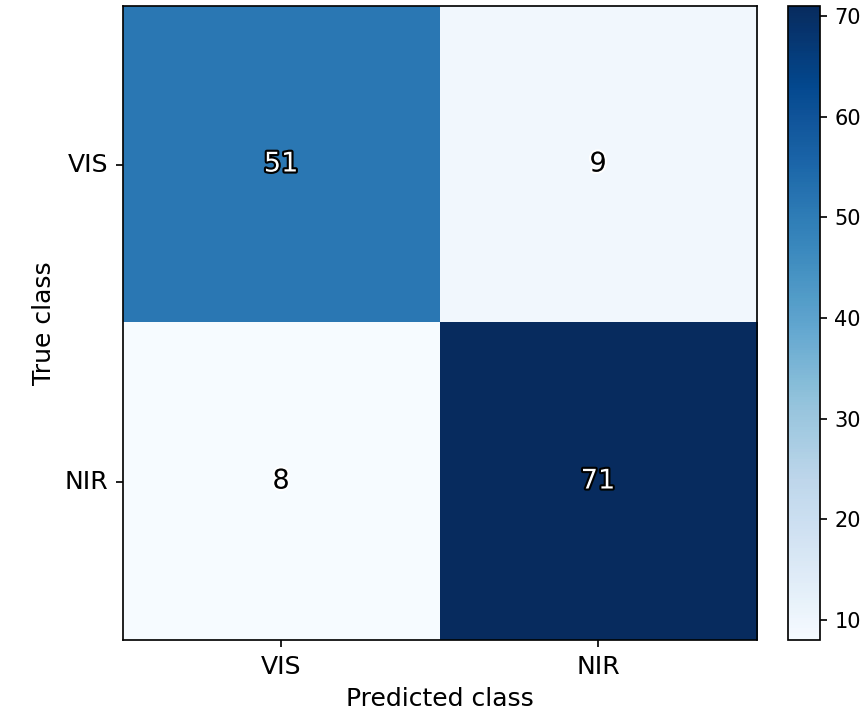}
            \caption[Test dataset :  augmented vectors]%
            {{\small Test dataset : 140 4D (augmented) vectors}} 
            \label{fig:}
        \end{subfigure}
        \caption[]
        {Confusion matrices for the binary classification of polymers in the dataset comprising 557 species.~CPU results (on left), QPU results (on right).~Each polymer
is represented by a 4D feature vector encoding its monomeric chemical structure.~They are randomly sampled from the larger datasets described in Sec.~\ref{sec:comp_details}
 and employed in Fig.~\ref{fig:CM_results_large_data_set}.} 
        \label{fig:Confusion_matrices_small_data_set}
    \end{figure*}
also demonstrates effective classification of VIS and NIR polymers, as evidenced by the confusion matrices: false positives and false negatives are an order of magnitude smaller for both training and test datasets (Fig.~\ref{fig:CM_results_large_data_set}).~Our analysis also reveals a somewhat superior performance of the hybrid model when utilizing  augmented input 4D vectors (see results for pair (5281, $\checkmark$) in Table~\ref{tab:results}).~This difference, however, is attributed to the distinct DNN feature extractors used for generating the 4D feature vectors with or without augmentation (see Secs.~\ref{sec:DNN_extractor_theory} and~\ref{sec:comp_details}).~Finally, note that the predictions for both train and test sets have relatively small error bars as computed from the Poisson distribution.

We subsequently analyzed a subset of 557 polymers, randomly sampled from the original dataset of 5281 data points.~Careful consideration
\begin{figure*}
    \centering
    \includegraphics[scale=0.5]{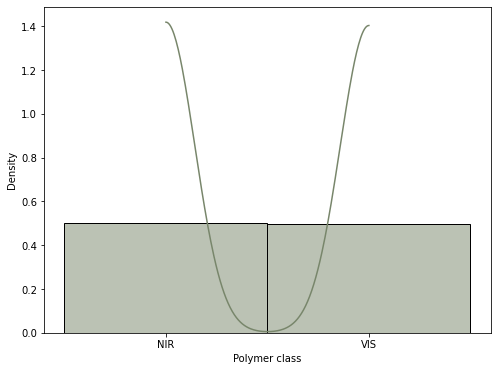}
    \caption{Density distribution and a normalized count of the polymer classes in the dataset of 557 polymers.~The dataset was sampled out randomly of the dataset of 5281 species.}
\label{fig:Distributions_small_data_set}
\end{figure*}
was given to ensure this smaller subset remained representative of the larger dataset (Fig.~\ref{fig:2D_input_data}) and balanced with respect to the VIS and NIR classes (Fig.~\ref{fig:Distributions_small_data_set}).~This dataset was designed to enable the quantum experiments on Quandela's QPU Ascella. 
\begin{table}[ht]
    \caption{Train and test accuracies from perfect and noisy simulations on a CPU using either strong ideal SLOS simulation backend or a SLOS backend with a noisy source, respectively.~Results from the quantum experiment carried out on Quandela's QPU Ascella ($\texttt{qpu:ascella}$) are shown in the last row of the table. Input data consists of the 4D (augmented) vectors.}
    \centering
    \begin{tabular}{c|c|c}
    \toprule
         &Train accuracy     & Test accuracy \\ \hline 
        $\texttt{ideal SLOS}$  &0.86 & 0.86 \\
                $\texttt{noisy SLOS}$  &0.88 & 0.83 \\
        $\texttt{qpu:ascella}$ & 0.87 &0.88 \\ 
        \hline
    \end{tabular}
    \label{tab:results2}
\end{table}

Exploiting this dataset, we have carried out three sets of experiments: (1) perfect simulations using strong ideal SLOS simulation backend on a CPU (2) noisy simulations on a CPU with a SLOS backend and a noisy source and (3) quantum experiments on Quandela's QPU Ascella ($\texttt{qpu:ascella}$).

Consistent with our approach for the larger dataset, we have first reported results on mean training and test accuracies derived from 5 noisy simulation runs using the SLOS backend on a CPU.~While comparable to the outcomes from the larger dataset, the mean test accuracy from the runs with the augmented 4D vectors is somewhat lower, albeit with similar standard deviations (see second (5281, $\checkmark$) and fourth (557, $\checkmark$) rows  in Table \ref{tab:results}).~Regarding the 4D numerical representations obtained directly (no augmentation), the standard deviation is approximately an order of magnitude larger for the smaller dataset (see first (5281, $\times$) and third (557, $\times$) rows in Table \ref{tab:results}).~Consequently, direct comparison of results between the small and large datasets is not straightforward.

We report next in table \ref{tab:results2} on the outcomes of perfect and noisy simulations on a CPU, as well as quantum experiments for both the training and test datasets.~Figure~\ref{fig:Confusion_matrices_small_data_set} shows results from the noisy simulations and quantum experiments.~Note, that all reported values are derived from a single computational run (thus, no standard deviations are provided in the table).~Efficient training of the circuit on the QPU remains challenging and hence, no statistically significant averages could be obtained.~Despite these challenges, an efficient training of the circuit in a single computational run was achieved.~The outcome of this run is presented in the final row of Table \ref{tab:results2}.~Regarding the CPU-based noisy SLOS simulations, we selected the best result (presented in the second row of Table \ref{tab:results2}) from the 5 computational trials  based on test accuracy metrics.~This selection facilitates a more direct comparison with the quantum device's outcome.~Note, also that we have used the 4D (augmented) vectors in this set of experiments. 

The CPU noisy simulation yielded lower test accuracy on the smaller dataset compared to the larger one, as previously noted.~Analysis of the confusion matrices in Fig.~\ref{fig:Confusion_matrices_small_data_set} reveals that the trained model exhibits lower performance for one of the two classes.~Interestingly, the quantum experiment on Ascella yields accuracies for both training and test datasets comparable to those obtained from CPU noisy simulations on the larger 4D (augmented) dataset.~Furthermore, it surpasses the performance of the trained model from CPU noisy simulations on the smaller dataset.

\section{Limitations and Outlook on Scalability}

At the current hardware scale, the number of available (fault-free) qubits and the allowable circuit depth are too limited to yield any form of quantum advantage for chemistry-native classification tasks.~In this regime, classical machine learning methods are fully capable of solving the problem with comparable or superior performance.~Accordingly, the results reported here should not be interpreted as evidence of quantum advantage.~Instead, the role of the quantum processing unit in this work is to validate feasibility and to explore hybrid quantum-classical workflows under realistic hardware constraints.

On the other hand, the proposed framework highlights a physically motivated scaling pathway.~As quantum hardware advances to support larger numbers of low-noise qubits, deeper circuits (as defined by the number of quantum gates), and improved gate fidelities, it will become possible to embed higher-dimensional input data directly into quantum circuits.~In such a regime, it becomes conceivable to operate directly on the original 139-dimensional polymer descriptors, thereby eliminating the need for classical neural-network-based pre-processing and dimensionality reduction. 

In the present study, we adopt a transfer learning strategy in which a classical neural network is used to compress the original high-dimensional polymer descriptors into a lower-dimensional representation and thus, to bridge the gap between high-dimensional chemical data and the constraints of near-term QPUs.~This hybrid approach enables us to leverage current QPUs while preserving a meaningful connection to realistic, high-dimensional chemical datasets.~It serves as a proof of principle that QPUs can be integrated into realistic materials-informatics pipelines, even in the absence of quantum advantage.

As QPU capabilities expand in terms of qubit number, gate fidelity, and circuit depth, the evolution will progressively shift the balance of this hybrid architecture toward the quantum component.~As these advances materialize, the reliance on classical preprocessing can be reduced or removed altogether, enabling quantum models to assume a more central role in feature learning and classification.~In this sense, the present results establish a baseline for future investigations at larger scales, where the question of quantum advantage can be meaningfully revisited.

\section{Conclusions}\label{sec:conclusions}
Recent research efforts \cite{ref42, Ascella_2023} have laid the groundwork for photonic-native QNN classifiers and their implementation on photonic chips.~While these studies were pivotal in establishing the feasibility of photonic QNNs, they were limited to relatively simple, toy datasets.~More precisely, the authors of Ref.~\cite{ref42} explored the concept through simulations on linearly-separable, circle, and moon dataset, while in reference~\cite{Ascella_2023} experiments were conducted using the iris dataset. 

In this work, we build upon these ideas, extending their application to a real dataset of polymers.~We have tested the performance of a photonic-native QNN for the task of polymers gap classification  assessing thus, its practical utility.~We have thereby combined the quantum model with a classical deep neural network, leveraging the concept of transfer learning.~This hybrid model aims to harness the strengths of both quantum and classical computing paradigms.~We evaluated our model's performance through both quantum simulations and a proof-of-principle run on a QPU.~The results demonstrate reasonable accuracy, suggesting the potential of this hybrid approach for real-world classification tasks.

Polymers present an ideal playground for quantum simulations and experiments due to their nature as stochastic mixtures of macromolecules with complex chemical and topological structures.~QNN algorithms can leverage the large Hilbert space to encode compactly the complex macro-molecular structures.~While we expect that advancements in quantum hardware are still necessary to achieve a quantum utility in the fields of chemistry or materials science, the application of NISQ algorithms to relevant datasets like ours remains highly valuable as it provides  insights into the potential and limitations of quantum computing in real-world scenarios. 

\begin{acknowledgments}
We are grateful to Quandela for providing their cloud-based quantum computer for the proof-of-principle run in this work.~And we are grateful to Arno Ricou for providing technical support and running the computational pipeline on Ascella.~We are also grateful to Alexia Salavrakos for proofreading the manuscript and for her valuable suggestions and comments that enhanced its quality.
 
\end{acknowledgments}

\bibliographystyle{plainnat}
\bibliography{polymer_qpu}

\appendix
\section{Polymer data description}\label{sec:polymer_data}
\begin{figure*}[!htb]
    \centering
\includegraphics[width=1.7\columnwidth]{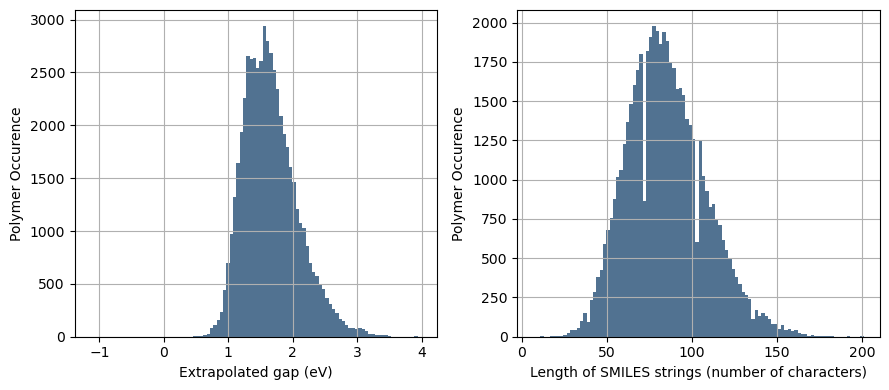}
    \caption{Dataset statistics: (left) distribution of the polymer extrapolated gaps (eV) (right) distribution of the length of the monomers' SMILES strings as measured by the number of characters.}
    \label{fig:data_statistics}
\end{figure*}

\begin{figure*}[!htbp]
    \centering
\includegraphics[width=1.8\columnwidth]{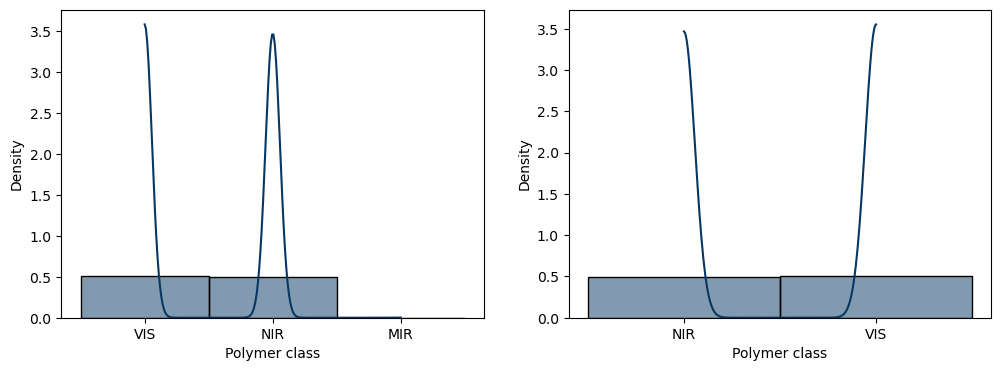}
    \caption{Dataset statistics: Density distributions and normalized counts of the polymer classes in the datasets of (right) 54335 and (left) 52815 polymers.}
    \label{fig:distribution_classes_big_dataset}
\end{figure*}

The polymer dataset yield by the B3LYP-DFT calculations \cite{ref4} contains 54335 unique polymers.~Each polymer is represented by the canonical SMILES of its monomer and the extrapolated value of the polymer gap.~Figure \ref{fig:data_statistics} shows the distributions of the extrapolated
values of the gaps and the SMILES strings lengths, respectively in the dataset.~Following the authors of Ref.~\cite{ref41}, we
 have classified the polymer gaps into three classes as a function of the spectroscopic energy ranges:~VIS, NIR and mid-infrared (MIR) range (see Table \ref{tab:polymer_classes}).~This 
\begin{table}[!htb]
    \centering
    \begin{tabular}{c|c}
    Polymer class & Gap value ranges \\ \hline
         NIR&  (0.4, 1.6]\\
         VIS& (1.6, 4.0]\\
         MIR& [0.025, 0.4] \\ \hline
    \end{tabular}
    \caption{Polymer classes according to extrapolated gap values.}
    \label{tab:polymer_classes}
\end{table}classification relies on the fact that the dataset contains multi-ring heterocycle compounds, conjugated polymers used in OPV-applications.~As shown in Fig.~\ref{fig:distribution_classes_big_dataset},
the distribution of polymer classes reveals a predominance of the NIR and VIS classes within the dataset.~The latter exhibits a balanced representation between these two classes.~The few MIR polymers are discarded here, thereby reducing the polymer analysis to a binary classification task. 

Returning to the distribution of the SMILES strings length in the dataset (Fig.~\ref{fig:data_statistics} (on right)), we note that it is symmetric and non-uniform (see Table \ref{tab:smiles_distribution}), 
\begin{table}[!htbp]
    \centering
    \begin{tabular}{c|c}
    Characteristics & Values \\ \hline
         mean&  85\\
         median& 83\\
         skewness & 0.46\\
         SMILES maximum length &  201 \\ \hline
    \end{tabular}
    \caption{Characteristics of the distribution of the SMILES strings length in the dataset.}
    \label{tab:smiles_distribution}
\end{table}
with 98 \% of the SMILES strings having less than 140 characters.~To encode the polymers (see, appendix \ref{sec:Encode_smiles}), we selected a critical value for the length of the SMILES strings, $L_c$, of 140 characters.~SMILES with $L_c < 140$ were padded with zeros, whereas those with $L_c > 140$ were discarded, thereby equalizing the length for all SMILES retained.~After identifying~\footnote{We used the standard deviation method on the normalized gaps distribution as well as boxplot visualisations.} and removing samples non-representative~\footnote{samples which gap value differs significantly from those of the majority of samples} for our dataset,
we obtain a dataset of 52815 polymers and their extrapolated gaps.~The polymer classes distribution remains the same after the dataset transformations (Fig.~\ref{fig:distribution_classes_big_dataset} (left)).

\section{Encoding the monomer SMILES}\label{sec:Encode_smiles}
Each monomer SMILES is encoded
following the strategy from Refs.~\cite{ref5} and \cite{Polymer_paper}.~First, we have constructed a dictionary of the unique characters in the SMILES corpus of the dataset.~Each character is encoded by an integer reflecting its position in the dictionary.~The positions range from 0 to the total number of unique characters minus 1 (0-based indexing).~For our dataset this dictionary contains 34 unique characters:
\begin{lstlisting}[language=python]
    {'#':0, '%':1, '(':2, ')':3, '-':4, 
    '.': 5, '/': 6, '0': 7, '1': 8, 
    '2': 9, '3': 10, '4': 11, '5': 12, 
    '6': 13, '7': 14, '8': 15, '9': 16, 
    '=': 17, '@': 18, 'C': 19, 'F': 20, 
    'H': 21, 'N': 22, 'O': 23, 'P': 24, 
    'S': 25, '[': 26, '\\': 27, ']': 28, 
    'c': 29, 'i': 30, 'n': 31, 'o': 32,
    's': 33}
\end{lstlisting}
Each SMILES is transformed into an integer-value vector with components being the index numbers of the SMILES characters in the dictionary.~For example, for the first three SMILES in the dataset, the corresponding vectors are 
\begin{align*}
v_1 = [19, 29,  8, ...,  0,  0,  0] \\
v_2 = [19, 26, 25, ...,  0,  0,  0] \\
v_3 = [19, 31,  8, ...,  0,  0,  0]
\end{align*}
signifying the following SMILES strings Cc1..., C[S..., Cn1..., respectively. The zeroes at the end have no special meaning and are only used to pad the sequences to a fixed length of 139 characters.

\section{Computational details on the classical DNN model}\label{sec:DNN_extractor}
\begin{figure*}[!htbp]
    \centering
\includegraphics[scale=0.55]{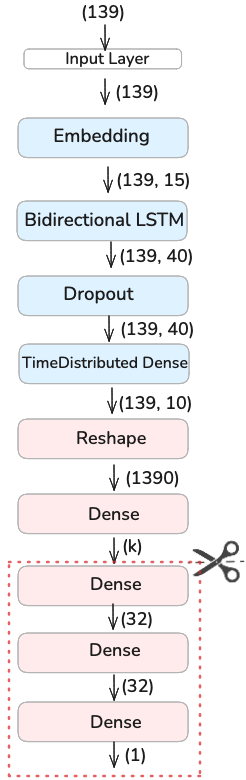}
    \caption{Classical DNN (from top to bottom: first layer [InputLayer] $\xrightarrow{}$ last layer [Dense]) Each layer is depicted by a rectangular containing information about the layer's type (Input, Embedding, LSTM, Dropout, Dense, ...) and the data dimensions at the input and output of the layer. For example, (None, 139, 40) allows a variable batch size (denoted by None), each sample in the batch having the shape (139, 40). The classical DNN feature extractor is obtained by discarding the last 3 Dense layers of the trained DNN model.}
    \label{fig:lstm}
\end{figure*}

The classical DNN architecture (see Fig.~\ref{fig:lstm}) employed in this study was already described in our previous work on photonic quantum computing for polymer classification \cite{Polymer_paper}.~For the sake of clarity, we recall it briefly.  

The architecture bears
close resemblance to classical RNN-LSTM neural networks, but it contains supplementary dense layers.~The first layer is an embedding layer that is a dense representation in which similar chemical species have similar encodings.~It is followed by conventional Bidirectional RNN-LSTM and Time Distributed Dense layers.~We then added three Dense layers, followed by a fourth Dense layer serving as the DNN output.~A dropout layer with a default value of 0.0 is also added for a possible regularization of the network.~We did not explicitly optimize two hyperparameters for the current task: the number of LSTM units in the Bidirectional LSTM layer-which also determines the number of units in the TimeDistributed Dense layer-and the dimensionality of the embedding layer output.~Their values were set to 20 and 15, respectively.~These values were found to be optimal in a regression study on the extrapolated gaps of the same dataset, using the same LSTM layers.

ReLu activation functions were used for the second and third Dense layers, whereas the first Dense layer was equipped with a tanh activation function.~This choice   
ensures that the derived  polymer vector components remain within the range [-1., 1.].~The fourth Dense layer (output) was supplied with a sigmoid activation.
      
The classical DNN feature extractor is then obtained by discarding the last three Dense layers of the trained DNN.~Applied to the dataset of 5281 polymers, it yields the  data vectors $\mathbf{x}'$ shown in Fig.~\ref{fig:feature_pipeline}, one per polymer, and depicted by the matrix on the right side:
\[
\begin{bmatrix}
[19,  0, 19, ...,  0,  0,  0],\\
[19, 19,  2, ...,  0,  0,  0],\\
[19, 23, 19, ...,  0,  0,  0],\\
..., \\ 
[19, 23, 29, ...,  0,  0,  0],\\
[19, 22,  8, ...,  0,  0,  0],\\
[19, 31,  8, ...,  0,  0,  0]
\end{bmatrix} 
\longrightarrow 
\begin{bmatrix}
       [-0.52132058,  0.292934],\\
       [ 0.84487766, -0.65577406 ], \\
       [ -0.05386608,  0.63849235 ], \\
       ..., \\
       [ 0.98356616, -0.61915606], \\
       [ 0.92054486, -0.5835743 ], \\
       [0.29776308,  0.39558443] \\
\end{bmatrix}
\] 
This is an example of the dimension reduction obtained with the classical DNN feature extractor.~The compact 2D vectors (on right) represent the polymer's chemical structure and its relation with a given class.~The dataset
    \begin{figure*}[!htpb]
        \centering
        \begin{subfigure}[b]{0.5\textwidth}
            \centering
            \includegraphics[width=\textwidth]{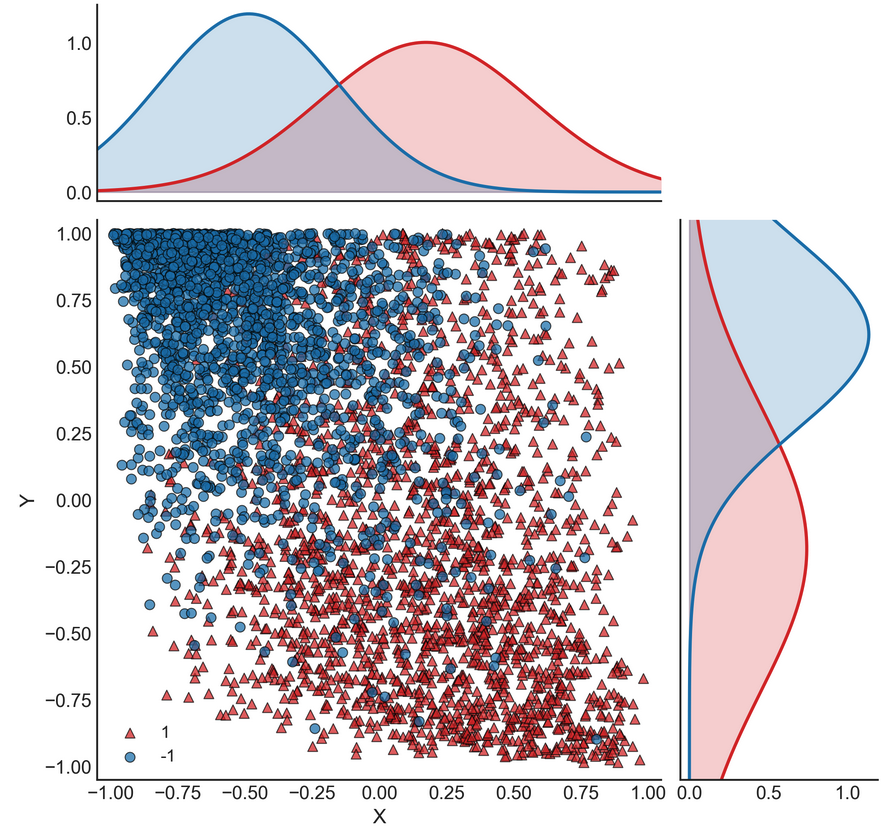}
            \caption[Train set]%
            {{\small Train dataset : 3960 2D vectors}}    
            \label{fig:}
        \end{subfigure}
        \hfill
        \begin{subfigure}[b]{0.49\textwidth}   
            \centering 
            \includegraphics[width=\textwidth]{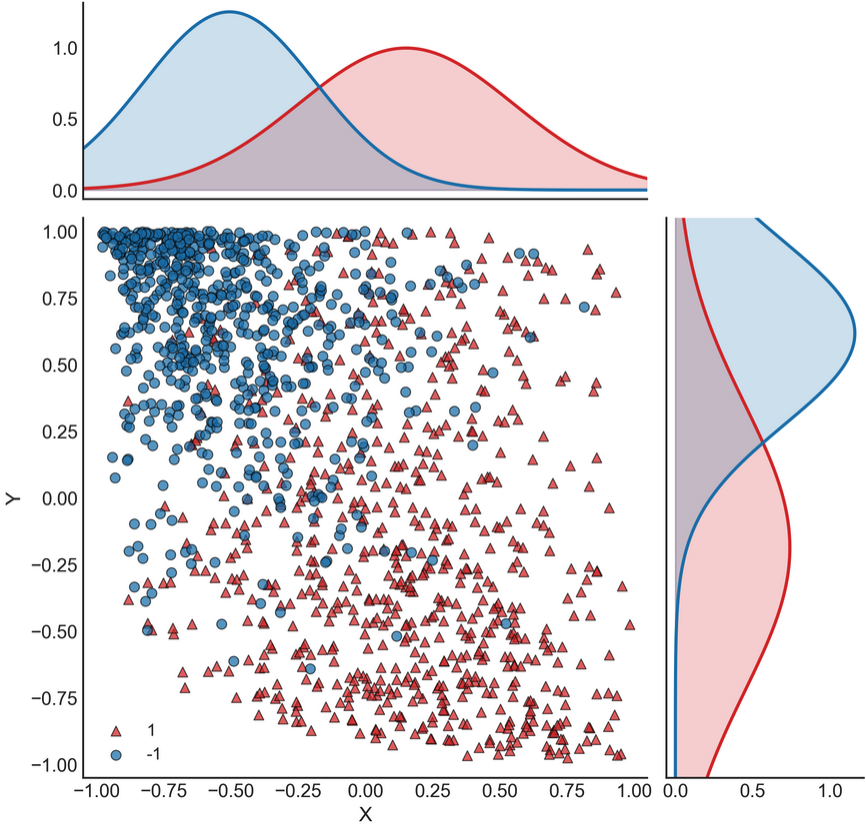}
            \caption[]%
            {{\small Testing dataset: 1321 2D vectors}} 
            \label{fig:}
        \end{subfigure}
        \caption[Illustration of the polymer 2D dataset used as an input data to the linear QPC after being augmented to 4D.]
        {Distribution of the polymer feature vectors yield by the classical DNN feature extractor with k = 2.~The central panel displays the coordinates 
of individual feature vectors in the learned feature space, with each point representing one data sample.~Two classes are shown: class NIR or –1 (blue circles) and class VIS or +1 (red triangles).~The top and right panels show the marginal distributions of the feature coordinates  for each class.~The marginals are Gaussian probability density functions computed from the empirical mean and standard deviation of each feature coordinate within a given class.~The shaded regions indicate the corresponding Gaussian densities and highlight the degree of overlap between the two classes along each dimension.~The feature vectors are used as an input data (only for CPU simulations) to the linear quantum photonic circuit after being augmented to k = 4.} 
        \label{fig:2D_input_data_big_data_set}
    \end{figure*}
of 2D vectors is next divided into a train 
and test subsets in the ratio (75:25) after being augmented to their 4D counterparts.~The  2D vectors of both subsets are shown in Fig.~\ref{fig:2D_input_data_big_data_set}. 
\begin{figure*}[!htpb]
        \centering
        \begin{subfigure}[b]{0.5\textwidth}
            \centering
            \includegraphics[width=\textwidth]{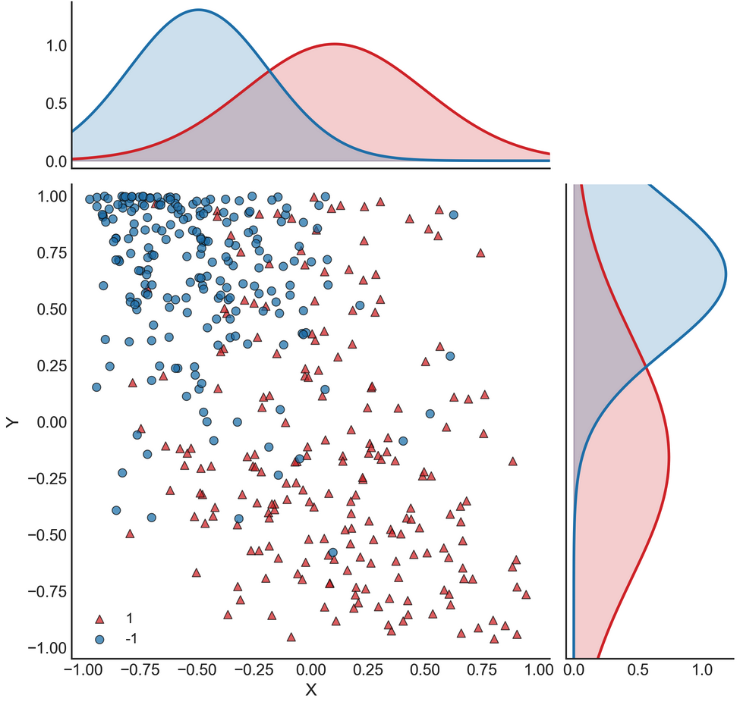}
            \caption[Train set]%
            {{\small Train dataset : 417 2D vectors}}    
            \label{fig:}
        \end{subfigure}
        \hfill
        \begin{subfigure}[b]{0.49\textwidth}   
            \centering 
            \includegraphics[width=\textwidth]{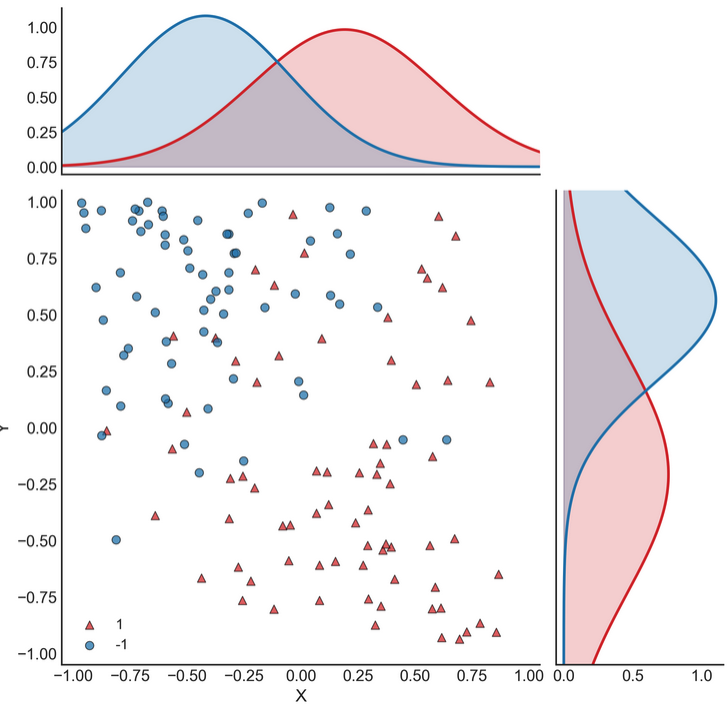}
            \caption[]%
            {{\small Test dataset: 140 2D vectors}} 
            \label{fig:}
        \end{subfigure}
        \caption[Illustration of the polymer feature vectors yield by the DNN feature extractor with k = 2.~They are used as an input data to the linear QPC.]
        {Distribution of the polymer feature vectors yield by the DNN feature extractor with k = 2.~Two classes are shown: class NIR or –1 (blue circles) and class VIS or +1 (red triangles).~The central panel shows the feature space populated by samples from the two classes, while the top and right panels display Gaussian marginal distributions of the individual feature vector coordinates for each class.~The data points are sampled out from the datasets in Fig.~\ref{fig:2D_input_data_big_data_set} and used (in both CPU experiments and in the proof-of-principle QPU ) as an input data to the linear quantum photonic circuit after being augmented to k = 4.  } 
        \label{fig:2D_input_data}
    \end{figure*}

\end{document}